\documentstyle[twoside,fleqn,espcrc2,epsf]{article}

\title{Analysis of finite temperature phase transition using level spacing
\thanks{Work supported in part by US DOE grants 
DE-FG0201ER41165 (HN) , DE-AC02-98CH10886 (FB), 
NSF grant PHY-0300065 (RN) and Jefferson Lab (RN).}}
\author{F.~Berruto\address{Physics Department, Brookhaven National Laboratory, Upton, NY 11973, USA}, R.~Narayanan\address{Department of Physics, Florida International University, Miami, FL 33199, USA}, H.~Neuberger\address{Rutgers University, Department of Physics and Astronomy, Piscataway, NJ 08855, USA}\\}

\begin{document}

\begin{abstract}
Let $B$ be the largest spacing between adjacent eigenvalues of the Polyakov
loop. We propose to employ the distribution of $B$ as an order parameter for the
finite temperature phase transition in $SU(N)$ lattice gauge theories. Using
smeared links to reduce ultraviolet fluctuations, we carry out a test
for the gauge group $SU(3)$.
\vspace{0mm}
\end{abstract}

\maketitle

The lattice regulator is a powerful non-perturbative framework to study 
the thermal properties of gauge theories. Polyakov loops provide 
a useful order parameter for the study of the finite temperature 
confinement-deconfinement phase transition of non-Abelian $SU(N)$ 
gauge theories. 
Polyakov loops are 
not properly renormalized quantities and in four 
dimensions are strongly affected 
by ultra-violet fluctuations. It is well known that in the 
continuum limit at fixed temperature the expectation value 
of an unrenormalized Polyakov loop vanishes even in the deconfined phase, due 
to divergent self-energy contributions to the free energy.
\begin{figure}[htb]
\epsfxsize=\hsize
\epsfbox{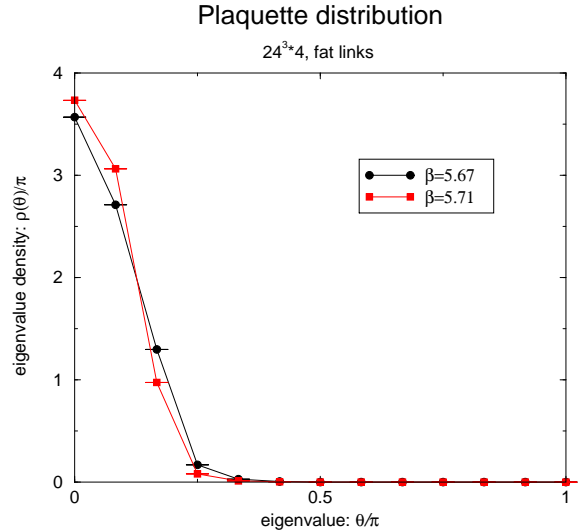}
\vspace{-12mm}
\caption{\it{Eigenvalue distribution of the plaquette in
the confined and deconfined phase.}}
\vspace{-7mm}
\label{plaq}
\end{figure}

For $N\le 3$, a lattice theory with Wilson action undergoes a bulk cross-over, 
separating the lattice strong- and weak-coupling regimes~\cite{creutz1}. 
The bulk cross-over is characterized by a steep increase in the average 
plaquette value and occurs at a fixed lattice coupling $\beta_B$. 
This steep increase occurs when the probability of $1\times 1$
Wilson loops to have eigenvalues near $-1$ decreases dramatically 
(for $N=\infty$ a true gap opens up). 
For $N\ge 5$ the cross-over becomes a discontinuity and there is a phase transition, but no symmetry breaks and there is no continuum counterpart
to this transition.  However, the bulk cross-over can affect the finite temperature transition~\cite{wingate} on coarse lattices.

The Polyakov loop $L$ is the parallel transporter around a closed loop in the 
``temperature'' direction and gets multiplied by a $Z_N$ phase under
certain action preserving changes of link variables in the path integral. 
In the confined regime (the disordered phase of a gauge theory) the 
eigenvalues of the Polyakov loops are randomly distributed over the unit 
circle. At some critical $\beta_c$  the $Z_N$ symmetry 
gets broken and the eigenvalues of the Polyakov loop favor one of the 
$N$ roots of unity on the unit circle. $\beta_c$ scales with $N_t$, 
the length of the lattice in the temperature direction. In order 
to describe the physical 
finite temperature transition
it is necessary 
to have $\beta_c>\beta_B$. In our $SU(3)$ simulations 
the plaquette distribution already had  
an apparent gap for $N_t=4$ over the whole range of $\beta$ that we studied
as seen in Fig.~\ref{plaq}.

Let the gauge invariant eigenvalues 
of a given Polyakov loop $L$ be 
$e^{i\theta_j}\ ( j=1,2,\ldots,N)$, $-\pi\le\theta_1<\theta_2<\ldots<
\theta_N<\pi$. The maximal level spacing 
$B$ is defined as 
$
\mbox{max}\left\{ 2\pi -\theta_N+\theta_1,\ \theta_{N}-\theta_{N-1},
\ \ldots ,\ \theta_2-\theta_1 \right\}.
$      
\begin{figure}[htb]
\epsfxsize=\hsize
\epsfbox{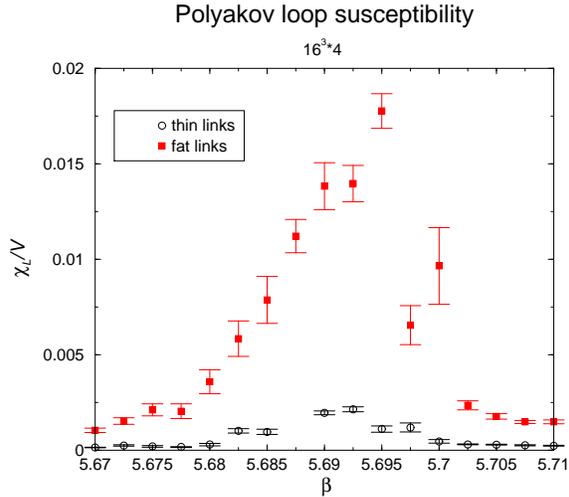}
\vspace{-12mm}
\caption{\it{Effect of APE smearing the links.}}
\vspace{-7mm}
\label{thin-fat}
\end{figure}
\begin{figure}[htb]
\epsfxsize=\hsize
\epsfbox{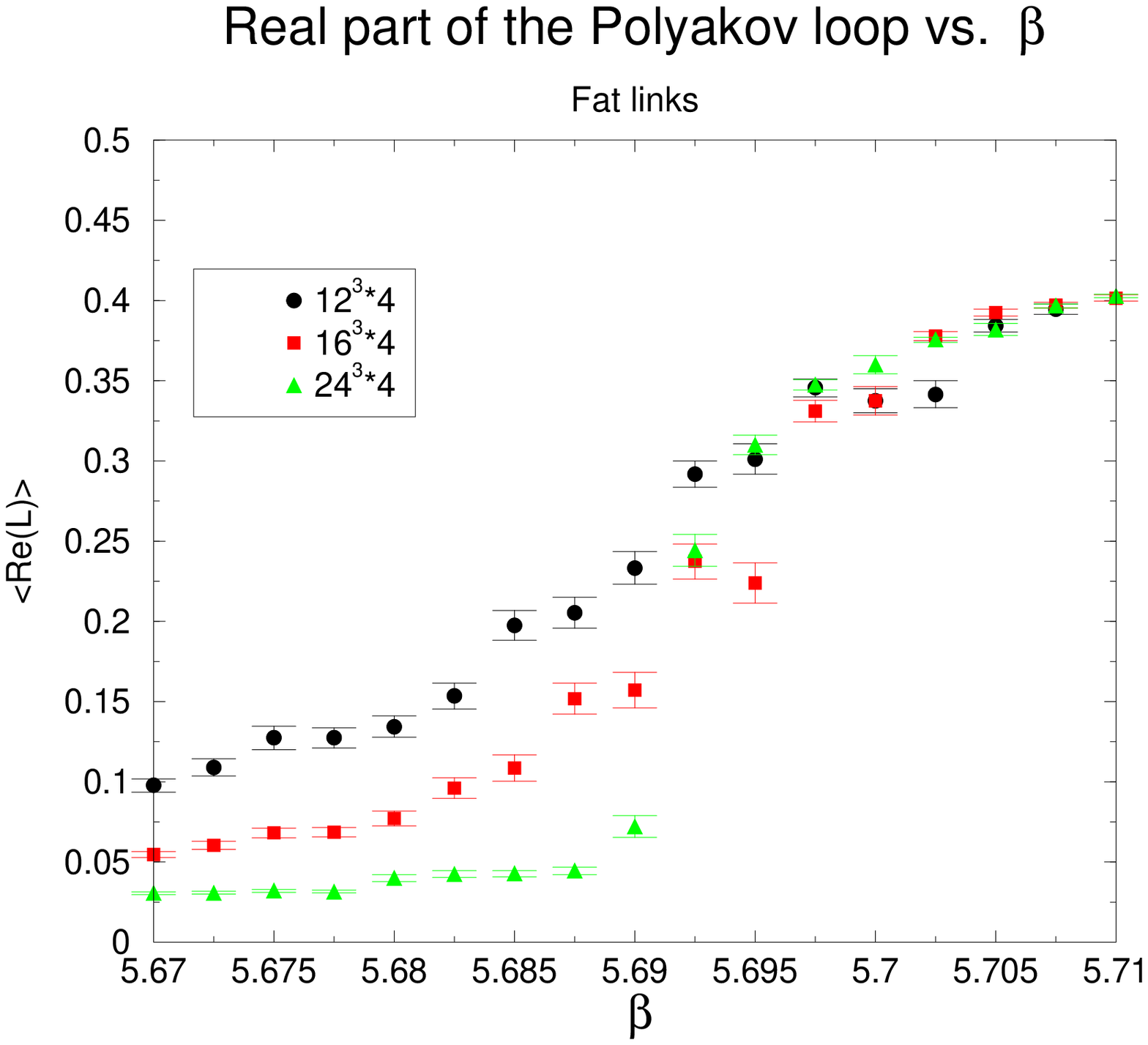}
\epsfxsize=\hsize
\epsfbox{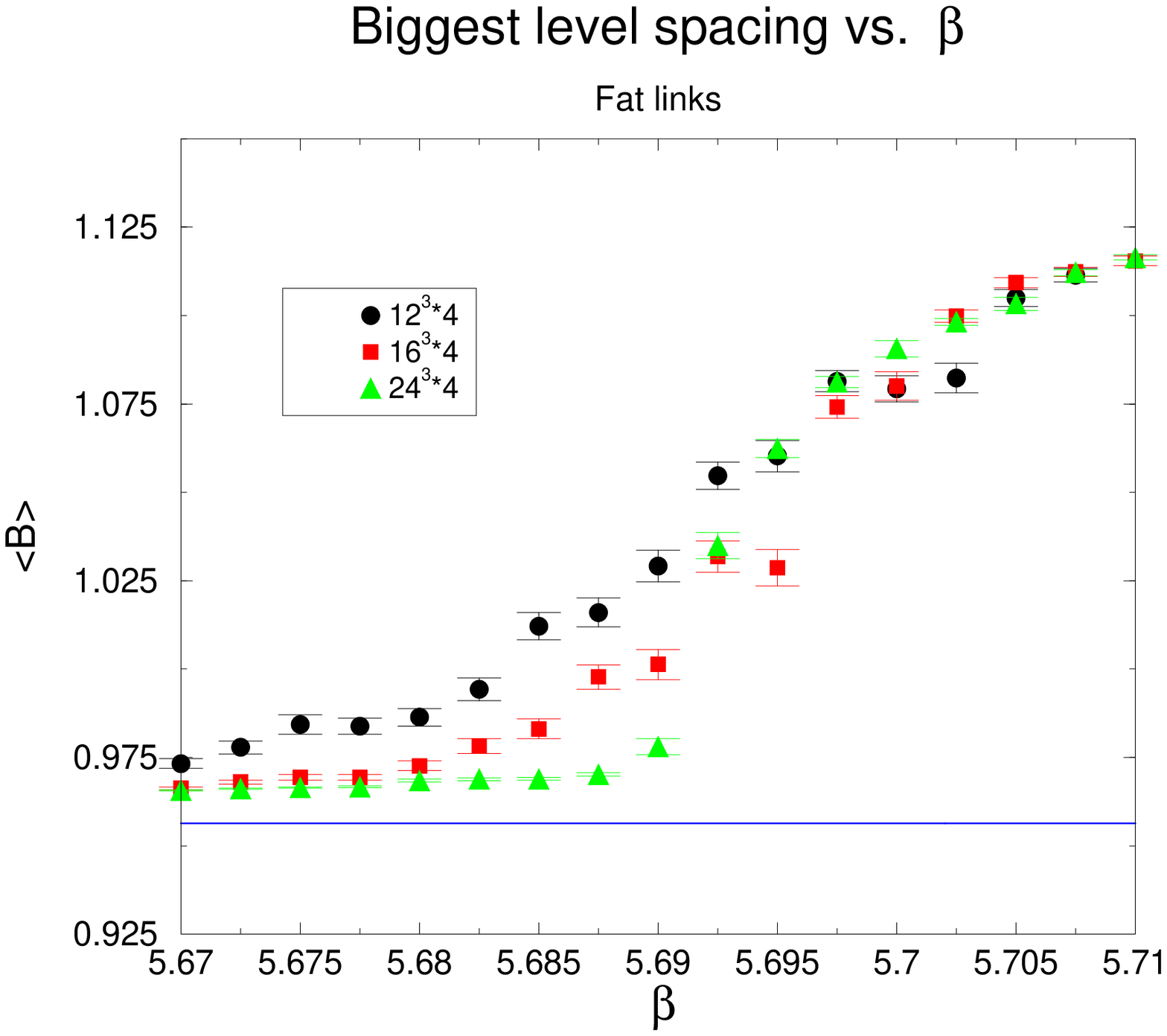}
\vspace{-12mm}
\caption{\it{$\langle \mbox{Re}(L)\rangle$ and $\langle B\rangle$ vs. $\beta$}}
\vspace{-7mm}
\label{lvsb}
\end{figure}
 
The distribution of $B$ changes through the 
finite temperature transition. It is neither affected by the particular
location on the unit circle where the spectrum condenses, 
nor by tunneling between the different $Z_N$ vacua. 
If the eigenvalues $\theta_j$ are randomly 
distributed on the unit circle $B$ will be of order $\sim 2\pi/N$; 
when the Polyakov loops get ordered, the concentration of angles at
one location creates an effective gap, increasing $B$. 
Random matrix theory can be used to describe the distribution of $B$
in the confined phase where it has one peak value. This is no longer
true in the deconfined phase where the distribution tends to have two
distinct peaks. The development of two peaks in the deconfined phase
is signaled by the single peak distribution developing a more populated
tail as one goes through the phase transition.
Deep in the confined phase one 
can compute the average value of $B$ by averaging over a very large number 
of random matrices. In the case of $SU(3)$ we get 
$\langle B_{\mbox{random}}\rangle=0.9564(1)$ and 
$\langle B_{\mbox{random}}\rangle\rightarrow 0$ for $N\rightarrow\infty$ in the case of  $SU(N)$. 
       
Following~\cite{fukugita}, we denote by $\mbox{Re}(L)$ the 
projection of the trace of $L$ onto the nearest $Z_3$ axis. 
The critical $\beta_c$ is generally determined by 
locating the peak in 
$
\chi_L=\langle\mbox{Re}(L)\mbox{Re}(L)\rangle-
\langle\mbox{Re}(L)\rangle \langle\mbox{Re}(L)\rangle\ .
$ Projection onto the nearest $Z_N$ works reasonably well
for $N=3$ but can become a problem close to the transition for
larger $N$. The observable $B$ does not suffer from such a problem
since it is defined as the biggest level spacing without any reference
to its location on the unit circle. 
\begin{figure}[htb]
\epsfxsize=\hsize
\epsfbox{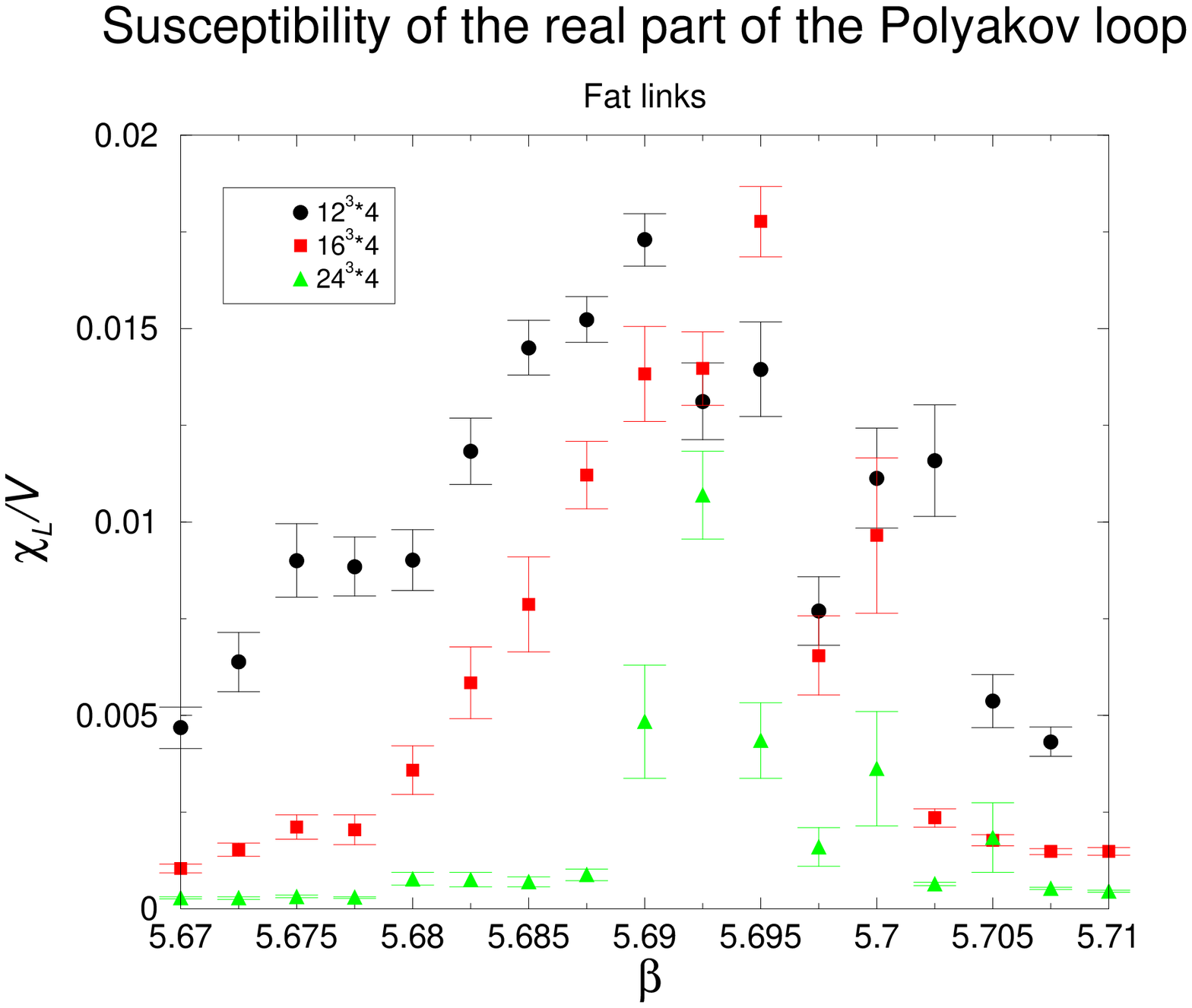}
\epsfxsize=\hsize
\epsfbox{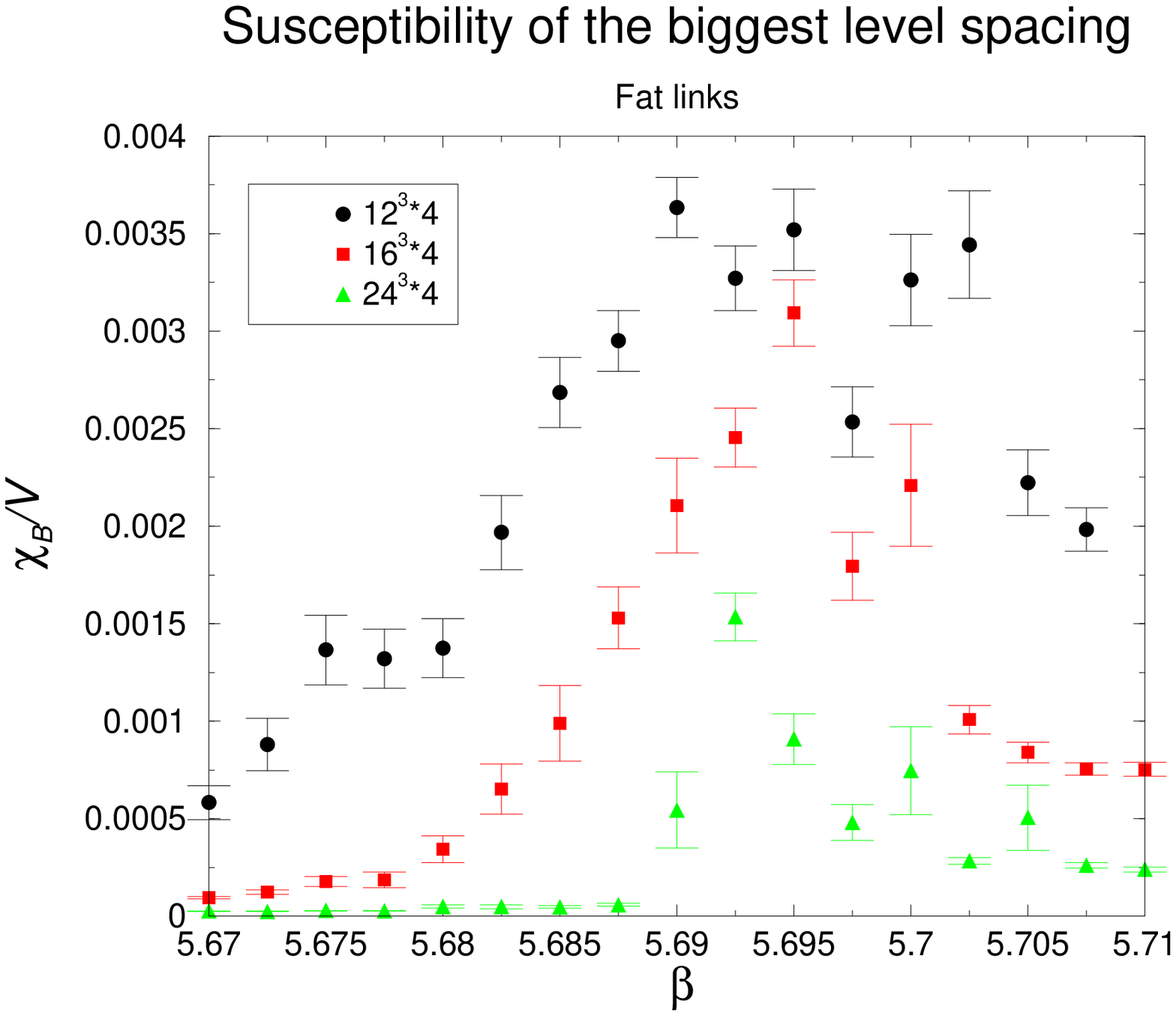}
\vspace{-12mm}
\caption{\it{Susceptibilities of $\langle \mbox{Re}(L)\rangle$ and $\langle B\rangle$.}}
\vspace{-7mm}
\label{chib}
\end{figure}

We performed various Monte Carlo simulations to study the finite temperature 
transition using the standard Wilson gauge action. 
A generic $SU(N)$ gauge fields update was implemented, even though
the present results are restricted to $SU(3)$. We used the 
heat-bath algorithm following Creutz-Kennedy-Pendleton
~\cite{creutz2}, where the update of one link corresponds to 
updates by all the $N(N-1)/2$ $SU(2)$ subgroups. 
In our Markov chain a step between two measurements consisted 
of one heat-bath sweep plus one over-relaxation sweep to avoid 
being trapped in a locally metastable state. Each over-relaxation update was 
done on the full $SU(N)$ group~\cite{knn}.
We considered lattices with 
$N_t=4$ and $L=12,16,24$ and typically we collected at least 
$15,000$ measurements. The errors were estimated with the jack-knife method. 

In order to suppress ultra-violet fluctuations we used APE smeared links 
following~\cite{degrand}. We smeared both the time-like and the 
space-like links on the lattice in parallel and repeated the process 
$N_t$ times. Subsequently, we used the resulting links to 
compute the Polyakov loops. 
In Fig.~(\ref{thin-fat}) one can see that smeared links provide 
a largely enhanced signal in the Polyakov loop susceptibility.    

The biggest level spacing turned out to be a good order parameter showing 
transitions consistent with traces of Polyakov loops as one can see in 
Fig.~(\ref{lvsb}) (the horizontal line represents $\langle B_{\mbox{random}}\rangle$,
the limiting value as $\beta\rightarrow 0$). 
By studying $\chi_B$ in 
Fig.~(\ref{chib}) we infer $\beta_c=5.6925(25)$ and this is consistent
with~\cite{fukugita}. Both $L=16$ and $L=24$ show well defined peaks
in $\chi_B$ and our estimate of $\beta_c$ was done
 without resorting to 
re-weighting and by working at relatively small volumes.  
The lack of monotonicity seen in 
Fig.~(\ref{chib}) 
in the deconfined phase is due to the system 
being trapped into one of the $Z_3$ vacua for a long time 
during the measurement history.

\end{document}